\magnification=1200
\magnification=1200

\def\d{\delta}

\def\k{\kappa}\def\o{\omega}\def
\p{\pi}\def\s{\sigma}\def\t{\tau}
\def\x{\xi}

\def\D{\Delta}\def\F{\Phi}\def\G{\Gamma}\def\L{\Lambda}
\def\O{\Omega}

\def\de{\partial}
\def\id{\equiv}\def\mo{{-1}}

\def\ex{{\rm e}}

\def\tran{transformations }\def\coo{coordinates }

\def\pb{Poisson brackets }

\def\poi{Poincar\'e }

\def\lt{Lorentz transformations }
\def\tl{transformation law }
\def\wrt{with respect to }\def\ie{i.e.\ }
\def\eom{equations of motion }

\def\section#1{\bigskip\noindent{\bf#1}\smallskip}
\def\nota{\footnote{$^\dagger$}}

\def\PL#1{Phys.\ Lett.\ {\bf#1}}
\def\PRL#1{Phys.\ Rev.\ Lett.\ {\bf#1}}
\def\PR#1{Phys.\ Rev.\ {\bf#1}}

 \def\IJMP#1{Int.\ J. Mod.\ Phys.\ {\bf #1}}
\def\MPL#1{Mod.\ Phys.\ Lett.\ {\bf #1}} 

\def\AoP#1{Ann.\ Phys.\ {\bf#1}}
\def\grq#1{{\tt gr-qc/#1}}\def\hep#1{{\tt hep-th/#1}}

\def\ref#1{\medskip\everypar={\hangindent 2\parindent}#1}
\def\beginref{\begingroup
\bigskip
\centerline{\bf References}
\nobreak\noindent}
\def\endref{\par\endgroup}

\def\del{\left(1-{p_0\over\k}\right)}
\def\bv{{\bf v}}
\def\dlt{deformed Lorentz transformations }
\def\tls{transformation laws }
\def\epk{\ex^{p_0/\k}}\def\emk{\ex^{-p_0/\k}}
\def\pik{{p_1\over\k}}\def\del{1-{p_0^2\over\k^2}}

{\nopagenumbers
\line{}
\vskip80pt
\centerline{\bf Hamiltonian formalism and spacetime symmetries}
\centerline{\bf in generic DSR models}

\vskip60pt
\centerline{{\bf S. Mignemi}\nota{e-mail: smignemi@unica.it}}
\vskip10pt
\centerline {Dipartimento di Matematica, Universit\`a di Cagliari}
\centerline{viale Merello 92, 09123 Cagliari, Italy}
\smallskip
\centerline{and INFN, Sezione di Cagliari}

\vskip100pt
\centerline{\bf Abstract}
\vskip10pt
{\noindent
We study the structure of the phase space of generic models
of deformed special relativity which gives rise to a definition
of velocity consistent with the deformed Lorentz symmetry.
In this way we can also determine the laws of transformation of
spacetime coordinates.}
\vskip100pt
\vfil\eject}

\section{1. Introduction}

Recently, the idea that the symmetry group of spacetime at energy
close to the Planck scale could be a deformation of the \poi group has
been widely investigated [1-4].
This hypothesis is motivated by the observation that the Planck energy
$\k$, whose role is essential in the formulation of the theories of quantum
gravity, might be a fundamental constant of physics on the same ground as
the speed of light, and should therefore be left invariant by the
group of transformation of spacetime. This may be achieved by deforming
the \poi group in such a way that its action on momentum space leaves
the energy $\k$ invariant [3-4].

Unfortunately, this assumption is not sufficient to single out a
unique deformation of the \poi group, even if one introduces
further physical requirements, as for example the request that in the
low energy limit the deformation tends to zero. One is lead therefore to
define a large class of different models, usually called deformed
(or doubly) special relativity (DSR) theories. The first example of these,
obtained from purely algebraic investigations, was the $\k$-\poi group [1-2].
Later, different models were derived from physical arguments [3-4].

All these models are characterized by the property that the
deformations are realized as a nonlinear action of the
Lorentz group on the momentum space [4-5]. This definition
however leaves  the action of the Lorentz group on the coordinate
space undetermined. It is evident that such action cannot be the
same as in special relativity, and in particular cannot be
independent of the momentum of the particle on which it is applied.
A further complication is the possibility, suggested by the $\k$-\poi
approach, that the geometry of spacetime be noncommutative [1-2].

From these considerations, it is also clear that the kinematics and the
dynamics of point particles must be modified if one wants to obtain a
picture consistent with the deformed spacetime symmetries.
In particular, the definition of the velocity of a particle is
problematic in DSR models [6-11]. In absence of a definite description of
spacetime, the velocity of a particle must in fact be defined in terms of
its momentum, but since the dispersion relations are deformed in DSR,
several inequivalent prescriptions are possible.
For example, if one adopts the naive definition $\bv_i=p_i/p_0$, the
velocity of a particle depends on its mass
[9], and the speed of light is energy dependent. The same problems arise
if one defines the velocity as $\bv_i=\de p_0/\de p_i$, as proposed by some
authors [3,12].
These drawbacks can be overcome if one requires that the velocity
be a property of the reference frame rather than of a specific
object and hence defines it in terms of boosts [10].

The expression for the velocity obtained in this way can be derived
from a Hamiltonian description of the motion of free particles only by
postulating noncanonical \pb [2,7-8]: in particular the \pb between space
and time coordinates cannot vanish. This property can be interpreted as
the classical counterpart of a noncommutative geometry. Although
several specific examples are given in the literature [7-8,13-14], no
general prescription is known for defining the Hamiltonian structure for
generic DSR models.

Fixing the Hamiltonian structure is also useful for the determination
of the \tls of coordinates. In [13], in fact, the \tls were derived from
the requirement that the action functional be a scalar under deformed
Lorentz transformations (DLT).
Although the methods of [13] worked well for the Maguejo-Smolin model [4],
they led to inconsistencies in the case of the $\k$-\poi model of ref.\ [1].
In fact, the same definition of velocity can be obtained from several
inequivalent Hamiltonian structures, but in general these do not lead to
the correct transformation rules under DLT.
One must therefore check that the velocity transforms in the correct way.
This was not the case for the Hamiltonian structure of the $\k$-\poi model
discussed in ref.\ [13].

In this paper, we try to extend the results of [13] to generic DSR
models, giving the conditions that must be satisfied by the \pb in order
to obtain the definition of velocity of ref.\ [10] obeying the correct
transformation laws. Although in the general
case these conditions are too difficult to be solved, we give some
explicit examples where this can be done.

\section{2. Hamilton equations}
According to the approach of ref.\ [5], since the symmetry
group of DSR theories is a nonlinear realization of the Lorentz
group, there must exist a function $\p=\F(p)$, with inverse
$p=\bar\F(\p)$, that maps the physical momentum $p$
into an unphysical  momentum $\p$ which transforms linearly
under Lorentz transformations.

The action of a \dlt on $p$ will then be given by the composition
$$p'=\bar\F\circ\L\circ\F(p),\eqno(1)$$
where $\p'=\L(\p)$ is the linear action of a Lorentz \tran on
$\p$. The kinematical quantities of the physical theory
transforming in the correct way under \dlt should then be defined
through this mapping. For example, this method has been used to
obtain a consistent definition of the addition law for momenta [5].

Using this prescription, Kosinski and Maslanka [10] have shown that
a definition of the velocity vector compatible with the group
structure of the DLT is given by\footnote{$^\ddagger$}
{For simplicity, we work in two dimensions, but the results can be
easily generalized. We denote 2-vectors indices by $a,b...=0,1$, and
1-vectors by bold letters. The signature is $(+,-)$, and the \coo
of a particle are denoted by $q_a$.}.
$$\bv={\p_1\over\p_0}={\F(p_1)\over\F(p_0)},\eqno(2)$$
\ie the velocity of DSR must coincide with that defined in the
standard way from the unphysical momentum $\p$.
It follows that under DLT the velocity transforms as in special
relativity. In particular, under a boost,
$$\d\bv=1-\bv^2.\eqno(3)$$
In [9] it was also argued that the definition (2) is the
only one that satisfies the natural requirement that the velocity
of a particle be independent of its mass. Moreover, it
implies that the speed of light is energy independent and always
equal to 1.

In order to write down a Hamiltonian formalism in which (2) arises
as the natural definition of the velocity, $\bv=\dot q_1/\dot q_0$,
we examine in more detail the theory.
According to the previous considerations, we assume that the
components of the unphysical momentum $\p_a$ ($a=0,i$),
satisfying the mass-shell constraint $\p_0^2-\p_1^2=m^2$,
can be written in terms of the physical momentum $p_a$ as
$$\p_0=F(p_0,p_1^2),\qquad\p_1=G(p_0,p_1^2)p_1,\eqno(4)$$
We write the inverse relations as
$$p_0=\bar F(\p_0,\p_1^2),\qquad p_1=
\bar G(\p_0,\p_1^2)\p_1.\eqno(5)$$
In this notation, the definition (2) of velocity is
$$\bv={G(p_0,p_1)\over F(p_0,p_1)}\,p_1.\eqno(6)$$
In general, this expression of the velocity cannot be obtained from
the Hamiltonian formalism with canonical Poisson brackets.
Nevertheless, as we shall see, it can be recovered if one admits a more
general symplectic structure [7-8].

It must also be remarked that the \tls for the momenta can be derived
from (1). In fact,
$$p'_a=W_a(p),\eqno(7)$$
where
$$W_0=\bar F(\p'_0,\p'_1),\qquad W_1=\bar G(\p'_0,\p'_1).\eqno(8)$$

As usual, the Hamiltonian $H$ for a free particle  can be  defined
as the Casimir operator of the deformed algebra,
$$H={m\over2}={1\over2m}(\p_0^2-\p_1^2)={1\over2m}(F^2-G^2p_1^2).\eqno(9)$$
Then the velocity of a particle is by definition
$$\bv={\dot q_1\over\dot q_0}={\o_{10}\de H/\de p_0+\o_{11}\de H/\de p_1
\over\o_{00}\de H/\de p_0+\o_{01}\de H/\de p_1},\eqno(10)$$
where $\o_{ab}=\{q_a,p_b\}$, and $\dot q_a\id dq_a/d\t$ is the derivative
of the position coordinate $q_a$ with respect to the variable $\t$ that
parametrizes the trajectory. The second equality follows from the
Hamilton equations. In the following, we shall assume that the $\o_{ab}$
are functions of the momenta, but not of the coordinates. We also
postulate $\{p_a,p_b\}=0$.

Equating the expressions (6) and (10) for $\bv$, one can obtain an algebraic
relation between the $\o_{ab}$. This relation is not sufficient to fix them
uniquely. However, further constraints arise from the \tl of the velocity.
Consider first the infinitesimal \tls of the momenta arising from (7),
$$\d p_a\id\{J,p_a\}\id w_a(p),\eqno(11)$$
where $J$ is the generator of the deformed boosts. Inserting (11) into
the Jacobi identities
$$\{\{J,q_a\},p_b\} +\{\{q_a,p_b\},J\} +\{\{p_b,J\},q_a\} =0,\eqno(12)$$
one can derive the infinitesimal \tls of the \coo
$$\d q_a\id\{J,q_a\}=u_{ab}(p)\,q_b,\eqno(13)$$
where the functions $u_{ab}$ depend on $w_a$, $\o_{ab}$ and their
derivatives with respect to $p_a$.

Since $\dot p_a=0$, deriving (13) \wrt $\t$, one obtains that the
$\dot q_a$ transform as the $q_a$.
For consistency, the velocity defined by (10) must also transform as (3).
Hence,
$$\d\bv=\d\left({\dot q_1\over\dot q_0}\right)={\dot q_0u_{1a}\dot q_a-
\dot q_1u_{0a}\dot q_a\over\dot q_0^2}=1-{\dot q_1^2\over\dot q_0^2},
\eqno(14)$$
and therefore the $u_{ab}$ must satisfy
$$u_{01}=u_{10}=1\qquad u_{00}=u_{11}=f(p),\eqno(15)$$
for some function $f(p)$.
Substituting (15) in (12) and equating (6) and (10) one obtains a system
of one algebraic and four partial differential equations for the five
functions $\o_{ab}$ and $f$.

After solving them, from the Jacobi identities
$$\{\{q_a,q_b\},p_c\} + {\rm perms.} = 0.\eqno(16)$$
one can obtain the \pb between the coordinates. In general
$\{q_0,q_1\}\ne0$, indicating the necessity of a noncommutative geometry.

It is interesting to note that, if the conditions (15) hold, the line
element $d\s^2=dq_0^2-dq_1^2$ transforms in a simple way, as
$$\d(d\s^2)=2f(p)\,d\s^2,\eqno(17)$$
and is therefore possible to construct an invariant "metric" by
multiplying $d\s^2$ by a suitable function of the momentum $p_a$.

Moreover, following [8], we notice that (6) and (10) imply that
$$\dot q_0= {A(p)\over m}\,F(p),\qquad\dot q_1= {A(p)\over m}\,G(p)\,p_1,
\eqno(18)$$
for some function $A(p)$. In term of differentials,
$$dq_0= {A(p)\over m}\,F(p)\,d\t,\qquad dq_1= {A(p)\over m}\,G(p)\,p_1\,d\t,
\eqno(19)$$
and hence
$$dq_0^2-dq_1^2=A^2{(F^2-G^2p_1^2)\over m^2}\,d\t^2=A^2(p)\,d\t^2.\eqno(20)$$
Consequently, the proper time $d\t$ is given by the line element $d\s$ times
a function of
the momentum. It is easy to see that the proper time so defined must be
invariant under DLT and can then be identified with the above defined
"metric".
It must also be remarked that since in general the variable $\s$ is
different from
the proper time $\t$ that parametrizes the trajectories, the modulus of
the 2-velocity is not unitary. A more detailed discussion of the
interpretation of these results will be given elsewhere.

\section{3. Examples}
The conditions introduced in the previous section give rise to a system of
partial differential equations for the $\o_{ab}$, that in general is extremely
difficult to solve.
However, if $F$ and $G$ depend only on the energy $p_0$, it reduces to
a system of ordinary differential equations, and there is some chance to
obtain a solution.
For example, this is possible in the case of the Maguejo-Smolin model, as
discussed in [13]. We present here two other models where an explicit solution
can be obtained, namely the Poincar\'e subalgebra of the deformed conformal
algebra introduced by Herrantz in ref.\ [15], and the Heuson model of ref.\
[16].

\bigskip
\noindent{\it a) The Herrantz model}
\smallskip
\noindent{This model is defined by the functions [15]}
$$F=\k\left(\epk-1\right),\quad G=1,\eqno(21)$$
that give for the velocity
$$\bv={p_1\over\k(\epk-1)}.\eqno(22)$$
The \tls of the momentum under a boost of rapidity $\x$ are
$$\eqalignno{&p'_0=\k\log\D,\qquad
p'_1=p_1\cosh\x+\k(\epk-1)\sinh\x,\cr
&{\rm with}\qquad\D=1+(\epk-1)\cosh\x+\pik\sinh\x.&(23)}$$
In infinitesimal form,
$$\d p_0=p_1\emk,\quad\d p_1=\k(\epk-1).\eqno(24)$$

The Hamiltonian reads
$$H={1\over2m}\left[\k^2(\epk-1)^2-p_1^2\right],\eqno(25)$$
and the conditions of the previous section are satisfied by the Poisson
structure
$$\o_{01}=\o_{10}=0,\quad\o_{00}=1,\quad\o_{11}=-\epk.\eqno(26)$$
In view of (16), $\{q_0,q_1\}=q_1/\k$.
From (25) and (26) follow the Hamilton equations
$$\dot q_0={\k\over m}\,\epk(\epk-1),\qquad \dot q_1=\epk{p_1\over m},\eqno(27)$$
from which one recovers (22).

Using (12) one can then obtain the explicit infinitesimal \tls for the \coo
$$\d q_0=q_1+\pik\,\emk q_0,\qquad \d q_1=q_0+\pik\,\emk q_1.\eqno(28)$$
The line element $d\s^2=dq_0^2-dq_1^2$ transforms therefore as
$$\d(d\s^2)=2\pik\emk d\s^2,\eqno(29)$$
and hence the form
$$d\t^2=\ex^{-2p_0/\k}(dq_0^2-dq_1^2)\eqno(30)$$
is invariant under infinitesimal DLT. This is the proper time,
as defined in (20).

Following ref.\ [13], it is also possible to find the finite form of
the deformed \tls for the coordinates consistent with the Hamiltonian
structure. It is known that
the Hamilton equations for systems with nonstandard symplectic
structure can be derived from an action principle [18].
Given a phase space with symplectic structure
$\{Q_A,Q_B\}=\O_{AB}$, where $Q_A$ denotes either the \coo or
the momenta, one defines the functions $R^A(Q_A)$ such that
$${\de R^A\over \de Q_B}-{\de R^B\over \de Q_A}=\O^{AB},\eqno(31)$$
where $\O^{AB}$ is the inverse of $\O_{AB}$.
The Hamilton equations can then be obtained varying with
respect to $Q_A$ the action
$$I=\int(R^A\dot Q_A-H)d\t.\eqno(32)$$
Note that in general the action so defined contains derivatives
of the momenta.

In our case, we define $Q_1=q_0$, $Q_2=q_1$, $Q_3=p_0$, $Q_4=p_1$.
Inverting $\O_{AB}$, one finds for $\O^{AB}$ the nonvanishing
components $\O^{13}=-\O^{31}=-1$, $\O^{24}=-\O^{42}=\emk$,
and $\O^{34}=-\O^{43}=q_1\emk$.
Solving (31), one has then
$$R^1=p_0,\quad R^2=-p_1\emk,\quad R^3=-{p_1q_1\over\k}\,\emk,
\quad R^4=0.$$
Substituting in (32) and integrating by parts one obtains
$$I=-\int\left[q_0\dot p_0-q_1\emk\dot p_1+H\right]d\t,\eqno(33)$$
and can identify the variables conjugated to the momenta $p_a$ as
$$r_0=q_0,\quad r_1=-q_1\emk.\eqno(34)$$

In order for the action to be invariant under DLT,
the $r_a$ must transform controvariantly, \ie as
$$r'_a=\L_{ab}(p)\,r_b,\eqno(35)$$
where $\L_{ab}=(\de W_b/\de p_a)^\mo$.
From (34) and (35), it follows after some calculations that the \coo
transform as
$$q'_0=\emk\D(\cosh\x\ q_0+\sinh\x\ q_1), \quad
q'_1=\emk\D(\sinh\x\ q_0+\cosh\x\ q_1),\eqno(36)$$
\ie as a \lt times a momentum-dependent factor.

Differentiating (36), and recalling that $\dot p_a=0$ by the field equations,
one easily sees that the $\dot q_a$ transform as the $q_a$ and that the
velocity $\bv$ transforms in the required fashion. Moreover, the "metric"
(30) is invariant also under finite boosts.

\bigskip
\noindent{\it b) The Heuson model}
\smallskip
\noindent{This model was introduced in ref.\ [16] (see also [17]) and is analogous
to that of ref.\ [4]. It is defined by}
$$F={p_0\over\sqrt{\del}},\quad G={1\over\sqrt{\del}}.\eqno(37)$$
From (6) one gets the velocity
$$\bv={p_1\over p_0}.\eqno(38)$$
The \tls of the momentum under a boost of rapidity $\x$ are
$$\eqalignno{&p'_0={p_0\cosh\x+p_1\sinh\x\over\G},\qquad
p'_1={p_0\sinh\x+p_1\cosh\x\over\G},\cr
&{\rm with}\qquad\G=\sqrt{1-{p_0^2\over\k^2}+{1\over\k^2}(p_0\cosh\x+p_1\sinh\x)^2}.
&(39)}$$
In infinitesimal form,
$$\d p_0=p_1\left(\del\right),\quad\d p_1=p_0\left(1-{p_1^2\over\k^2}\right).
\eqno(40)$$
The Hamiltonian reads
$$H={1\over2m}\ {p_0^2-p_1^2\over\del},\eqno(41)$$
and the conditions of consistency are solved by the functions
$$\o_{01}=-{p_0p_1\over\k^2},\quad \o_{10}=0,\quad\o_{00}=\del,\quad\o_{11}=-1.
\eqno(42)$$
From (41) and (42) follow the Hamilton equations
$$\dot q_0={p_0/m\over\del},\qquad \dot q_1={p_1/m\over\del}.\eqno(43)$$
which yield the velocity (38).
The Jacobi identities (16) imply a nontrivial Poisson bracket between space and
time coordinates, $\{q_0,q_1\}=p_0q_1/\k^2$.

Using (12) one can deduce the infinitesimal \tls for the \coo
$$\d q_0=q_1+{p_0p_1\over\k^2}q_0,\qquad \d q_1=q_0+{p_0p_1\over\k^2}q_1,
\eqno(44)$$
and for the line element $d\s^2$,
$$\d(d\s^2)=2{p_0p_1\over\k^2}d\s^2.\eqno(45)$$
It follows that the form
$$d\t^2=\left(\del\right)(dq_0^2-dq_1^2)\eqno(46)$$
is invariant under infinitesimal DLT.

Proceeding as in the previous example, one can obtain also the finite
transformations. The \eom can be derived by the action
$$I=-\int\left[{q_0+p_0p_1q_1/\k^2\over1-p_0/\k^2}\,\dot p_0-q_1\dot p_1
+H\right]d\t,\eqno(47)$$
and therefore the variables conjugated to the momenta $p_a$ are
$$r_0={q_0+p_0p_1q_1/\k^2\over1-p_0/\k^2},\quad r_1=-q_1.\eqno(48)$$
From the transformation rules of the $r_a$, one can then obtain the
\tls for the \coo
$$q'_0=\G(\cosh\x\ q_0+\sinh\x\ q_1), \quad
q'_1=\G(\sinh\x\ q_0+\cosh\x\ q_1).\eqno(49)$$
As in the previous example, they look like the \lt except for a
momentum-dependent factor.
One can easily check that under (49) the velocity $\bv$ transforms in the
correct way and the "metric" (46) is invariant.

\section{4. Conclusions}
We have established the conditions that the Poisson structure of
DSR models must satisfy so that the Hamilton equations yield an
expression for the velocity of a particle which is consistent
with the DLT, and have given two explicit examples of their application.

We have obtained one algebraic and four differential equations for
five unknown functions, and this should be sufficient to fix uniquely the
Poisson structure.
However, a general algorithm to solve these conditions is not
available. In particular, we have not been able to find a solution
in the case of the $\k$-\poi model of ref. [1-2].

From the requirement of invariance of the action it is also possible
to deduce the laws of transformation of the \coo of a particle.
Their main peculiarity is the dependence on the momentum of the
particle, and this suggests that a consistent description of
spacetime in DSR theories should involve the full phase space.
In particular, the proper time, invariant under DLT, is given by the
product of the line element $d\s$ with a suitable function of the
momentum.

\vfill\eject

\beginref
\ref [1] J. Lukierski, A. Nowicki, H. Ruegg and V.N. Tolstoy, \PL{B264},
331 (1991); J. Lukierski, A. Nowicki and H. Ruegg, \PL{B293}, 344
(1992).

\ref [2] J. Lukierski, H. Ruegg and W.J. Zakrzewski, \AoP{243}, 90 (1995).

\ref [3] G. Amelino-Camelia, \IJMP{D11}, 35 (2002), \PL{B510}, 255 (2001).

\ref [4] J. Magueijo and L. Smolin, \PRL{88}, 190403 (2002).

\ref [5] S. Judes and  M. Visser, \PR{D68}, 045001 (2003).

\ref [6] T. Tamaki, T. Harada, U. Miyamoto and T. Torii,
\PR{D65}, 083003 (2002); \PR{D66}, 105003 (2002).

\ref [7] J. Lukierski and A. Nowicki, Acta Phys. Polon. {\bf B33}, 2537
(2002).

\ref [8] A. Granik, \hep{0207113}.

\ref [9] S. Mignemi, \PL{A316}, 173 (2003).

\ref [10] P. Kosi\'nski and P. Ma\'slanka, \PR{D68}, 067702 (2003).

\ref [11] M. Daskiewicz, K. Imilkowska, J. Kowalski-Glikman, \PL{A323},
345 (2004).

\ref [12] G. Amelino-Camelia, F. D'Andrea and G. Mandanaci, JCAP
{\bf0309}, 006 (2003).

\ref [13] S. Mignemi, \PR{D68}, 065029 (2003).

\ref [14] J. Kowalski-Glikman, \MPL{A17}, 1 (2002).

\ref [15] F.J. Herranz, \PL{B543}, 89 (2002).

\ref [16] C. Heuson, \grq{0305015}.

\ref [17] D. Kimberly, J. Magueijo and J. Medeiros, \PR{D70}, 084007
(2004).

\ref [18] R.M. Santilli, {\it Foundations of theoretical mechanics II},
Springer-Verlag, New York, 1983.

\endref
\end